\documentstyle[11pt,aaspp4]{article}

\def\msun{{\ifmmode M_\odot \else {$M_{\odot}$}\fi}}
\def\lsun{{\ifmmode L_\odot \else {$L_{\odot}$}\fi}}
\def\sub#1{\ifmmode _{#1} \else $_{#1}$\fi}
\def\sup#1{\ifmmode ^{#1} \else $^{#1}$\fi}

\def\h2{}  

\lefthead{Hearn \& Lamb}
\righthead{Arp~119: A High Speed Galaxy Collision}

\begin{document}

\title{Arp~119: A High Speed Galaxy Collision with Episodic Star Formation}

\author{Nathan C. Hearn\altaffilmark{1} and Susan A. Lamb\altaffilmark{2, 3}}
\affil{Center for Theoretical Astrophysics,\\ Department of Physics, University of Illinois at Urbana-Champaign}

\altaffiltext{1}{Loomis Laboratory of Physics, 1110 W. Green Street, Urbana, IL 61801, USA. Email: n-hearn@astro.uiuc.edu}

\altaffiltext{2}{Loomis Laboratory of Physics, 1110 W. Green Street, Urbana, IL 61801, USA. Email: slamb@astro.uiuc.edu}

\altaffiltext{3}{also Department of Astronomy, University of Illinois at Urbana-Champaign}


\begin{abstract}
Colliding galaxies are excellent laboratories for studying galactic evolution and global star formation.  Computer simulations of galaxy collisions, in which at least one galaxy has a significant gaseous component, show the production of density enhancements and shock waves in the interstellar medium.  These high-density regions coincide with the locations of recent, large-scale star formation in observations of some real colliding galaxies.  Thus, combined n-body/hydrodynamic computer simulations can be used to explore the history and conditions of star forming regions in colliding galaxies.  We compare multi-wavelength observations of the Arp~119 system with a combined n-body/SPH simulation of colliding galaxies.  Most of the observations used here are gleaned from the literature.  Additionally, we obtained new near infrared (J- and H-band) observations of this system, using the NIRIM camera at the Mount Laguna observatory.  This new data adds information about the underlying, old stellar population.   

Arp~119 (CPG 29) is comprised of a southern member, Arp~119S (Mrk 984),
that has an extremely disturbed appearance, and a northern member, Arp
119N, a gas-poor elliptical.  The morphology of both members can be
fit well by a simulation in which a gas-rich disk galaxy has been impacted by an equal mass elliptical that had a trajectory approximately perpendicular to the plane of the disk, and passed through the disk slightly off-center. From our comparison, we find that the progression of recent
large-scale star formation in this galaxy can be accounted for by {\it a single \/}outwardly propagating collision-induced density wave in
the gas. We deduce that the star formation rate in this density wave was not a smooth function of time but that starbursts occurred episodically at intervals of approximately 25 to 30 Myr. The fit of the simulations to the
observations indicates that the collision between the two galaxies occurred
approximately 71 Myr ago.  At the current, projected separation of 53 kpc
(assuming $H_0 = 75$\,\hbox{km~s\sup{-1}~Mpc\sup{-1}}), we obtain a current relative space velocity between the two galaxies of approximately 850 km~s\sup{-1}.  This is strong evidence that the collision partner was Arp~119N, and not some
currently more distant galaxy.  Furthermore, the high relative velocity of
the pair, and the paucity of gas currently to be found in the elliptical, may
explain the very high velocity gas observed in the greatly disturbed disk
galaxy, Arp~119S.

\end{abstract}



\section{Introduction}

Collisions between galaxies cause fundamental changes in galactic structure, initiating large-scale star formation, and often triggering gas flow towards nuclear regions.  The stellar component of a galaxy becomes reorganized due to the changing gravitational potential during the collision.  The perturbations produce distinct patterns that last for at least a dynamical time, which is on the order of $10^8$ years for an L* galaxy. The effect on the interstellar gas is even more extreme because of the collisional nature of the gas.  Sharp peaks in density occur when flows of gas intersect due to orbit crossing (see Struck 1997 and references therein).  It is within these regions that strong shocks are found in hydrodynamic simulations and bursts of star formation are observed in real disk galaxies.  The effects of the collision may be very long lasting.

Collisionally produced ``ring'' galaxies and related objects are particularly interesting and useful in the study of collision-induced, global star formation because of their relatively simple geometry.  This type of ring galaxy is formed from a disk galaxy after a collision with another galaxy in which a significant component of the relative motion of the two galaxies is parallel to the disk's rotation axis.  (The second galaxy can be of any type, but its mass must be at least 10\% of the disk galaxy's mass if the results of the impact are to be readily observable.)  A collision of this kind produces flows of material moving toward and away from the impact point, with density waves in both the stars and the gas emanating away from this location.  These waves can take the form of complete rings, in the case of a nearly on-axis collision, or long arcs or strong arms when the impact parameter is larger (see Gerber \& Lamb 1994).  Such collisions also produce motions perpendicular to the plane of the disk, which eventually results in a displacement of the nucleus from the rest of the disk in the direction of motion of the second galaxy.  An off-center collision can cause the nucleus to be shifted along the plane of the disk, with respect to the outer disk material, towards the disk impact point.  (``Displaced'' nuclei are observed in some impacted galaxies, for example, Few \& Madore 1986.)

Arp~119, also known as CPG 29, is a pair of galaxies whose southern member (Arp~119S; Mrk 984) has a strongly disturbed, asymmetric form.  The visible- and B-band image of Arp~119S in Figure 1a (taken from Arp 1966) exhibits some of the remarkable features of this galaxy, including: a bright, circular ring surrounding the nucleus; a luminous arc north of the nucleus that extends to the west; a long arc of knots along the southern edge; and a ``veil'' of low-luminosity material in the east.  The distance to Arp~119 from Earth is 193 Mpc, and the two members are separated by roughly 53 kpc in the plane of the sky. (The value of assuming $H_0$ is assumed to be 75~\hbox{km~s\sup{-1}~Mpc\sup{-1}} throughout.) 

Observations taken at multiple wavelengths can be used to determine the locations of recent star formation, and the history of the collision-induced gas density wave in Arp~119S.  The H$\alpha$ and B-band images of Marziani, et al.\ (1994) highlight those regions where hot, massive stars have been formed.  Thus, the morphology in these bands provides information about the location of the collision-induced gas density wave in the recent past.  The CO (1--0) observations of Gao (1996) show where the current regions of cold, dense gas are located. These molecular gas regions can be directly related to dense structures in the simulated gas disk.  Thus, this set of observations can be used to infer how the gas density wave propagated through the disk of Arp~119S during the time since the collision.

Conversely, near-infrared observations reveal the distribution of the older (pre-collision) stellar population at the current epoch.  Like the gas, the arrangement of these stars is strongly affected by collisions.  However, the large-scale structures induced in the stellar distribution are not as sharply defined as those in the gas because the stars are non-collisional.  We have made new observations of Arp~119 in the J- and H-bands using the NIRIM infrared camera at the Mount Laguna Observatory, and we present these observations here.

Computational models generated with combined n-body/hydrodynamic techniques are used to study these collisional galaxy systems. A series of generic collision simulations covering an specific region of the parameter space were employed. In these simulations, a disk galaxy collided with an elliptical, which was initially on a trajectory perpendicular to the disk. Variations in both the impact parameter and the mass ratio of the two galaxies were included. We used three-dimensional visualization techniques to explore the simulations, and to determine the relationship between the varied parameters and the collision induced, large-scale morphological changes in the disk galaxy.  

When a real galaxy is observed to display some of these distinct morphological features, it may be possible to find a close match in the set of simulations. The model match is made by viewing the simulations from many different angles over the course of many timesteps until the proper model and viewing angle is found.  These are determined by matching dense structures in the projected simulation gas disk to those seen in observations of recent star formation regions and molecular gas in the real disk galaxy. Similarly, the near-infrared images can be compared with the distribution of the original stellar population in the simulation.  The simulation timestep showing the same features as those seen in the real galaxy's CO emission is referred to as the ``current epoch'' or ``present time.''

Once a matching computational model is found, the dimensions of the simulated disk galaxy are scaled to the size and mass of the real galaxy.  The evolution of dense, gaseous structures in the simulation can then be analyzed and compared to the observed regions of recent star formation in the real galaxy.  Large-scale formations seen in the H$\alpha$ and B-band observations are compared to the simulated gas disk at times prior to the current epoch.  These regions of new stars can be correlated with locations of high gas density and strong shocks in the simulated gas density wave.  The times at which the various star formation episodes occurred can then be estimated from the simulation.  These times are measured with respect to the time at which the collision took place, which we define as that time when the elliptical galaxy's center of mass was coincident with the midplane of the disk galaxy.  By comparing the simulation's dynamics with those observed in the real galaxy, an understanding of the history of and the conditions influencing star formation can be obtained.  (An introduction to this method is presented in Lamb, Hearn, \& Gao 1998, in which the galaxy system Arp 118 is discussed and analyzed.)  

In Section 2, we present the observations we have used to compare with the simulations, including our new near-infrared images of Arp~119.  Section 3 provides an overview of the computational methods used to produce the collision simulations. The process of finding an appropriate numerical model to compare with the Arp~119 observations is described.  Having used the observations to determine the applicable physical dimensions of the simulation, timescales are computed from the simulation that can be compared to the multi-wavelength observations of the system.  The physical scaling allows a discussion of the spatial evolution of star-forming regions, which is given in Section 4.  The velocity structure of the model system is presented in Section 5, and is compared with observed values. The model allows us to interpret the observations in a new way, and to constrain the systemic velocity of the Arp~119, given the observations of the disk rim velocities. These results and some of their implications are further discussed in Section 6.  

\section{Observational Data}

\subsection{A Review}

Figure 1a is an image of the Arp~119 system taken in the visible (Arp 1966). The northern member is Arp~119N, and the peculiar galaxy to the south is Arp~119S.  While the former appears to be a normal elliptical, the southern member has proven to be quite remarkable.  Arp~119S displays a strongly disturbed morphology with a substantial level of asymmetry, featuring an off-center nucleus.  There is significant evidence for recent and current large-scale star formation at several locations in the disk.  Marziani, et al.\ (1994) observed some spiral structure in Arp~119S typical of a Sc galaxy, as well as the presence of two, or possibly three, rings of different radii in the disk.  Arp~119S has been classified as a LINER galaxy by Mazzarella, Bothun, \& Boroson (1991) and Bernl\"{o}hr (1993).  Marziani, et al.\ (1994) found that it has multiple regions with LINER-like spectra.  

The line-of-sight velocities for the two galaxies of the Arp~119 system have been measured simulateneously from long-slit spectra placed across the nuclei of both galaxies. Mazzarella \& Boroson (1993) found a systemic velocity for 
Arp~119S of 14,379 km~s\sup{-1} using visible spectra, but Marziani et al.\ (1994), using the sodium D lines, found a systemic velocity for Arp~119S of 14,200 km~s\sup{-1} and a velocity of 15,000 km~s\sup{-1} for Arp~119N. 
This relative line-of-sight velocity of approximately 800 km~s\sup{-1} is to be compared to that determined by Mazzarella \& Boroson (1993), which is only 170 km~s\sup{-1}. 

A detailed study of the spectra of the Arp~119 system, together with B-band and H$\alpha$ images of the southern member, was conducted by Marziani, et al.\ (1994).  They detected several distinct morphological structures in the images of Arp~119S (including rings, arcs, and knots), as well as possibly four separate (line-of-sight) velocity components in the spectra from the slit position oriented along the north-south line passing through the nuclei. Two of these components are dominant in the H$\alpha$ data: one is near the ambient, possibly original velocity of the disk galaxy, and the other is a high-velocity component, redshifted possibly as much as 1300 km~s\sup{-1} with respect to the ambient velocity.  In between these two extremes, a smaller amount of emission is seen at two intermediate redshifts.  Horellou, et al.\ (1995) noted that H~I spectra of the system exhibited much structure, with at least two strong emission peaks seen at 14,150 and 14,350 km~s\sup{-1} (probably associated with the ambient velocity component of Arp 119S), as well as much weaker emission at 14,700~km~s\sup{-1} that can be associated with a mass of $9 \times 10^8~\msun$. This higher velocity gas could not be localized, but the association with Arp~119 was clear.

Marziani, et al.\ (1994) computed masses for both members of Arp~119 using B-band mass-to-light ratios. The mass for Arp~119N was calculated using a ratio of 8, a value typical of elliptical galaxies (see Faber \& Jackson 1976). This mass agreed with the dynamical mass of $4.4 \times 10^{11} \msun$ (Marziani, et al.\ 1994). A mass-to-light ratio of 3 was chosen for Arp~119S, which is in the range of those for Sc (4.7) and Sdm (1.7) galaxies as listed by Faber \& Gallagher (1979).  This ratio yielded a mass of $3.7 \times 10^{11} \msun$ for Arp~119S, about twice as large as the value indicated by rotation curve measurements within a radius of 17.6 kpc.  This apparent discrepancy is understandable, given that this rotation curve is derived from the motion of material through which the arc-like density wave has already traveled.  Models for collision-induced arc- and ring-like density waves show a decrease in rotation speeds after the passage of the wave, as compared to the rotation rate at that radius originally (see Lamb, Hearn, \& Gerber 2000).  Additionally, the rotational velocity determinations are sensitive to the choice of the galactic systemic velocity.  In the case of Arp~119S, there is enough uncertainty in the value of this quantity to bring the direction of rotation into question.  Determinations of rotational velocities in collisional disk galaxies are uncertain due of possible bulk radial motions, which are not present in typical quiescent galaxies.  (These last two issues are discussed in Section 5.1).

\subsection{Prior Observations used for Comparison with Models}

Observational images of Arp~119 are available for a limited selection of other wavelengths, including B-band, H$\alpha$, and CO (1--0) emission.  While radio continuum images would be helpful for finding regions of enhanced supernova activity, this range of wavelengths available to us is sufficient for determining the locations of the most recent large-scale star formation.
As stellar populations age, the number of hot, blue stars diminishes faster than the number of cooler, redder stars.  Thus, B-band emission serves as a good indicator of recent star formation.  The B-band image (Figure 1b) obtained by Marziani, et al.\ (1994) shows the central 40'' by 40'' region of Arp~119S.  A number of interesting features appear in this image.  A ring with a 6''--7'' radius can be seen surrounding the nucleus, as well as a bright arc located to the north that extends to the west.  Two knots are prominent along the vertical line north of the nucleus.  The one closest to the nucleus is thought to be an H~II region.  Mazzarella \& Boroson (1993) measured its velocity as 15,172 km~s\sup{-1} (thereby associating it with the high-velocity material in this system), and its apparent B-magnitude as 19.98 (compared to 17.51 for the nucleus).  Marziani, et al.\ (1994) detected both ambient and high velocity components in H$\alpha$ spectra of this knot.  In contrast, they detected only the high velocity component in the northern-most knot.  This feature appears as a ``cusp'' in the optical image (Figure 1a), and has an apparent B magnitude of 21.10 (Mazzarella \& Boroson 1993).  Marziani, et al.\ (1994) also note the presence of a knot directly to the east of the nucleus.  This knot appears to be coincident with the ring structure observed in the B-band image.  

H$\alpha$ emission originates from regions of strong UV flux, as around hot, young stars.  Figure 1c shows the Marziani, et al.\ (1994) H$\alpha$ image, with contours overlaid.  The enhanced emission appears to be more confined to the central regions of Arp~119.  The blue ring of 6''--7'' radius has no counterpart here, and the knot east of the nucleus is missing.  The northern arc is still visible, but it is smaller and has a small extension to the east.  The knot just to the north of the nucleus is also present, with an emission line luminosity comparable to that of the nucleus, although the continuum is significantly weaker in the knot. While N~II/H$\alpha$ emission in this knot is consistent with a stellar photoionization source, Marziani, et al.\ (1994) noted that an additional mechanism, such as one associated with shocks, may be necessary to account for the observed strength of the O~I emission.  An enhancement in H$\alpha$ emission is seen in a region southeast of the nucleus that has no clear counterpart in the B-band image.  The H$\alpha$ luminosity for the nuclear region has an upper bound of about $3.5 \times 10^7~\lsun$ (Mazzarella, Bothun, \& Boroson 1991).  

Gao (1996; 1997) produced the CO (1--0) image shown in Figure 1d, which displays an area of about 50'' by 50'' centered on Arp~119S.  CO emission is used to trace cold molecular gas.  Strong CO emission was detected from structures more than 10 kpc away form the nucleus.  In fact, only about 30\% of the CO emission in Arp~119S comes from the nuclear region, while strong emission is seen in the arc-like structures.  The CO image shows a knot just north of the nuclear region, coincident with the knot seen in B-band and H$\alpha$.  A northern arc is observed, in this case with a significant extension to the east.  Enhanced emission is present southeast of the nucleus, in the same general region as emission in seen in the H$\alpha$ image.  There is no clear evidence for the 6''--7'' ring seen in the B-band, except for a couple possible knots.  Horellou, et al.\ (1995) found strong H~I, CO (1-0), and CO (2-1) lines at the velocity of the main body of Arp~119S, about 14,300 km~s\sup{-1}.  The emission levels correspond to a molecular hydrogen mass of $4 \times 10^9 \msun$, or about one-third of the H~I mass.  However, this same study was unable to find the high-velocity component (around 15,300 km~s\sup{-1}) in either H~I or CO, suggesting an upper limit on the high velocity H\sub{2} mass of $2.4 \times 10^9 \msun$.

Far-infrared emission is associated with regions of newly formed stars whose UV emission is processed through dust.  IRAS data has been used by at least three groups to determine the luminosity of far-infrared emission for the Arp~119 system, as well as the ratio of far-infrared to B-band luminosity.  The results of these investigations vary widely, and throw the usefulness of these quantities into some doubt.  Mazzarella, Bothun, \& Boroson (1991) found log $L_{FIR} = 11.01$, and their measurement of log $L_B = 10.67$ produced an $L_{FIR}/L_B$ ratio of 2.18, where all luminosities are measured in \lsun.  Bernl\"{o}hr (1993) determined $L_{FIR} = 4.90 \times 10^{10}$ and $L_{FIR} / L_B = 3.4$.  Horellou, et al.\ (1995) found $L_{FIR} = 7.15 \times 10^{10}$, which yielded $L_{FIR} / L_B = 0.61$ given a value of $L_B$ calculated from the Third Reference Catalog of Bright Galaxies.  Based on the $L_{FIR} / M(H_2)$ ratio of 17.6 (where $M(H_2)$ is in units of \msun), the last group of investigators noted that Arp~119S is a little more efficient at forming stars than other galaxies in their sample, although it is not close to being an ultraluminous galaxy.

Near-infrared emission traces the distribution of the older stellar population, and is quite insensitive to recent star formation.  Bushouse \& Stanford (1992) observed Arp~119 in the J-, H-, and K-bands using the IRIM infrared imager on the 1.3-meter telescope at the Kitt Peak National Observatory.  They were able to detect the general outline of Arp~119S and the approximate position of its nucleus.  This can be seen in the image displaying their J-band observations and contours (Figure 1 in their paper).  Their contours for Arp~119S are slightly asymmetric, revealing the low-density material to the southeast and the more-luminous material to the west.  Both nuclear and global near-infrared magnitudes were measured for Arp~119S.  The global magnitudes, obtained on ``non-photometric'' nights, were $J = 14.67$, $J - H = 0.79$, and $H - K = 0.42$.

\subsection{New Near Infrared Observations}

We observed Arp~119 in the near-infrared using the one-meter telescope at the Mount Laguna Observatory.  H- and J-band images were obtained with the Near-Infrared Imager (NIRIM) camera.  This camera uses a 256 x 256 Rockwell NICMOS3 Hg:Cd:Te array as a detector, with a read noise of approximately 55 electrons, and a dark current of less than one electron per second. The camera, filter, and optics are described in Meixner, Young Owl, \& Leach 1999.  The reimaging optics used yielded a pixel size of 1'', and the seeing for both observations was roughly 3'' (FWHM).  At Arp~119's distance from Earth, one arcsecond on the sky is equivalent to approximately one kiloparsec.  

Arp~119 was observed in J-band on the night of January 23, 1999.  The J-band filter used is centered at a wavelength of 1.257 microns, and has a width of 0.293 microns.  Individual images were taken by choosing an on-object location, and then recording three exposures for 60 seconds each.  These were then followed by exposures of the same duration on an empty field for two different off-object locations.  This ON-OFF pattern was then repeated with new, slightly different on- and off-object locations.  Our J-band data represents a total on-object exposure time of 28 minutes, and is displayed in Figure 2a, with contours representing a range of flux levels superimposed. The data were flat-field corrected using the median of all off-object exposures. Bad pixels were eliminated from the frames prior to registration and combination in a final mosaic image.  

The H-band image with contours is shown in Figure 2b.  It is the result of 4.5 minutes of on-object exposures, obtained in 30-second segments, taken on the night of January 27, 1999.  The H-band filter had a central wavelength of 1.649 microns, and a width of 0.313 microns. 

The J-band image (Figure 2a) resolves the details of the Arp~119 system better than the H-band image (Figure 2b), most likely because of its longer exposure time.  The extent of the low-density ``veil'' to the southeast is much clearer, and the northern knots of Arp~119S are more distinct.  The emission of point-like sources, like the isolated source to the northwest of Arp~119S, is better resolved. A broad, cusp-like structure is detected in the old stellar population along the northern edge, which appears to point towards the elliptical, and coincides with the sharp cusp observed in the blue and visible light.

\section{Model fit}

\subsection{Matching to a Real System}

The purpose of building models of any physical system is to produce a simple representation that contains enough of the essential elements, or behavior, of the system that it can be used to make predictions about the origin or evolution of this, and similar systems.  Generic features of the behavior of the whole or parts of the real system, in a variety of circumstances, can be understood in terms of the behavior of this model when placed in a similar simulated environment.  The nature of the model and the extent of the simulated environments is restricted by the tools available to construct them. With the advent of computers, numerical modeling and simulation have become very powerful tools, opening up areas of research into real, very complex systems, such as galaxies, their formation and evolution. 

Studies of collisions between galaxies form an important part of this general investigation of galaxies, and their role in the formation and evolution of structure in the Universe.  Since the inception of this field of study by Toomre \& Toomre (1972), the numerical studies of colliding galaxies have taken two, complementary forms. On the one hand, numerical studies of collisions between pairs of galaxies, over several of the most important collision parameters, such as relative mass, impact parameter, and relative velocity, provide a framework for understanding the variety of morphological and other effects that collisions can produce.  On the other hand, it is of great interest to attempt to explain the structure and behavior of a real system, to understand, for example, how a particularly disturbed disk galaxy came to have the properties that it does, and to learn from this one system something about the underlying physical processes at work. 

We used a previously constructed group of simulations (see Gerber 1993; Gerber, Lamb, \& Balsara 1996; and Lamb, Hearn, \& Gerber 2000) to find a good fit to one particular observed system, and thereby deduce information about the large-scale star formation that occurred after the collision. That is, we have not attempted to build a precise model fit to the Arp~119 system. Rather, we chose to study the Arp~119 system because, among other reasons, we found that it is a close fit to one of the existing numerical models. The process of finding a model fit and the criteria for deciding that one has a good fit are very important. There is no numerical algorithm for measuring the ``goodness of fit,'' so the fitting process depends upon considerable experience with both the observations and the simulations. We are aided in the fitting process by the occurrence of strong morphological features in impacted disk galaxies, which allow us to put strong constraints on the models and simulations.

When we attempt to find a model fit to a real system, we are looking for fits to the generic features of disk galaxies that result from collisions. The morphological features can include rings and long arcs spanning the diameter of the disk, as well as isolated, kiloparsec-sized accumulations of gas (see, for example, Gerber \& Lamb 1994).  Some of the features in real colliding galaxies may reflect the pre-existing morphology, such as spiral arms. However, disk features dependent upon the initial galaxy structure are generally of less significance than the features induced by strong collisions of comparable mass galaxies, as is the case in Arp~119 (see Gerber, Lamb, \& Balsara 1996).  Measured internal velocities of a galaxy can also put strong constraints on a model, as is demonstrated by Lamb, Hearn, \& Gao (1998) for the Arp 118 system, where a detailed example of this method of simulation-to-observation matching is presented. Such a procedure allowed Curir \& Filippi (1996) to find an n-body disk-elliptical collision simulation with some of the notable kinematical features of the Arp~119 system. 

The general dynamics, relative orientation of the galaxies, and morphology of the Arp~119 system suggest that it can be well represented by a disk-elliptical collision in which the interpenetrating collision trajectory was roughly parallel to the disk spin-axis. This situation is the one modeled in the available set of simulations. However, this restriction in the collision geometry does not place strong restrictions on the applicability of the simulations because a fairly large range of incidence angles about the normal should produce similar effects in the disk.  A review of the numerical collision investigations of Toomre \& Toomre (1972) suggests that results for incidence angles as large as 30-degrees from the disk's normal will have noticeable similarity to those of normal encounters. 
 
In the simulations, the initial velocity of the elliptical with respect to the disk galaxy was set large enough for the system to be barely unbound, prior to the effects of dynamical friction.  In other words, a slightly hyperbolic relative trajectory was used.  As will be explored and discussed later in this paper, the Arp~119 pair has a large internal relative velocity, possibly over 1000 km~s\sup{-1}, which implies a hyperbolic orbit. Moles, Sulentic, \& M\'{a}rquez (1997) suggest that such high-velocity interactions can occur when galaxies fall into compact groups, and may even be necessary to prevent the group members from rapidly merging together.  

The mass ratio of the two galaxies is of central importance to the outcome of the collision.  The mass ratio establishes the amplitude of density wave in the ISM and the stars, as well as the speed at which the disturbances propagate through the disk.  Gerber, Lamb, \& Balsara (1996) found significant differences between disk-elliptical collisions in which the galaxy masses were equal and collisions in which the elliptical had only one-quarter the mass of the disk galaxy.  The equal mass collision had a much more remarkable effect on the vertical structure of the disk, significantly displacing the nucleus out of the plane of the disk, and producing a vertical separation between inflowing and outgoing gas.

To find a fit to an observed system, the set of simulations can be explored using three-dimensional visualization tools, employing multiple viewing angles. The process reveals the dependence of the predominant morphological results on the collision parameters, allowing possible fits to real systems. By reviewing a variety of models at different simulation timesteps, over a range of viewing angles, it is sometimes possible to find a fit good to the level of resolution present in the data.

It is important to note that one is always restricted by resolution. The numerical models and the simulations have a space and time resolution that depends on the computational space and time available for the simulation, and on the numerical methodology employed. Similarly, the observations of a real system, with which these simulations might be compared, have a resolution that depends upon the wavelength regime and equipment used. There are many processes that occur in galaxies on scales too small to be modeled directly on currently available computers.  If these processes have a significant effect on the galaxy on scales above the resolution limit, then they must be included in the simulations when possible. Algorithms for these ``sub-grid'' processes are often hotly debated because their implementation is often one of taste rather than physics. It appears that the best approach to such possible contributions (for example, energy input from star formation in colliding galaxies) is to parameterize the effect, and to explore the consequences of varying this parameter on the results. Such an investigation reveals the sensitivity of the results to this parameter. Given that the physics of star formation is not yet understood, this method is about the best one can use at present; and, at a minimum, it quantifies the uncertainty in the resulting simulations due to this lack of knowledge. 

Thus, in producing numerical simulations of colliding galaxies, we first construct models of galaxies that include the essential ingredients for our particular studies, that is, those attributes which we think are the most important in terms of what can be directly observed in real galaxies, or inferred from a combination of observations. Here we must be guided by that which is currently known of the physical constituents and their behavior, as we do not wish to incorporate unnecessary assumptions, or construct models that include poorly understood physics. Once numerical models are available, they can be used for a variety of simulated collisions in which several of the collision parameters are varied. From these simulations it is sometimes possible to find models that fit a particular colliding galaxy pair well. 

\subsection{The Numerical Simulations}

The computational simulations used for this analysis were generated by Gerber
(1993) using a combined n-body/hydrodynamic code (see Balsara 1990) which
he had modified for galaxy collision applications.  Further discussion is
presented in Gerber, Lamb, \& Balsara (1996).  The method of Smoothed
Particle Hydrodynamics (SPH) was used to model the hydrodynamic behavior of
the disk's gaseous component.  The collisionless stars and dark matter,
together with the gas, contributed to the gravitational potential, which
was calculated using standard particle-mesh (PM) methods.  The hydrodynamic
evolution equations included the artificial viscosity formulated by Balsara
(1990), allowing shocks to be captured by the SPH particles.  The gas is
assumed to be isothermal in these simulations, and only adiabatic thermal
processes were included.

The model elliptical was constructed out of 10,000 equal-mass ``star
particles'' representing the collisionless stellar and dark matter
components.  A Richie-King distribution was used with appropriate velocity
dispersions.  The disk galaxy halo was initially constructed in the same
manner.  The gravitational potential due to an infinite, thin exponential
disk was gradually imposed on this halo.  Once the halo had relaxed, the
imposed disk potential was removed, and the actual disk was populated with
25,000 collisionless particles representing the stellar disk component and
22,500 SPH particles representing the disk gas.  The disk ``star
particles'' had equal masses, while the SPH particles were given
non-uniform masses as suggested in Balsara (1990).  The star particles were
assigned an in-plane velocity dispersion sufficient to stabilize the disk
(see Toomre 1964), whereas the SPH particles were initially given only an
azimuthal velocity due to the rotation of the disk. The mass of the disk
galaxy was divided up among the various components such that the halo had a
mass 2.5 times that of the combined stellar and gaseous disks, and the gas
mass comprised 10\% of the disk mass.  Both the stellar and gaseous disks
were given an exponential surface density that was cut off after 4.4 radial
scale lengths, coinciding with the radial cut-off of the Richie-King halo. 
 
The matter distribution for the halos of real galaxies is not known in
great detail, but it will influence the way in which the elliptical
scatters from the disk galaxy during the collision. However, the assumption
of a roughly isothermal spherical distribution of matter in the ``dark''
halos of field galaxies is the simplest, and thus the best for use in these
studies unless observational evidence for some other distribution is
found.  Similarly, the distribution of mass between the disk material and
the halo is not known precisely.  The 1~:~2.5 mass ratio chosen for these
simulations gives a sufficiently massive halo to stabilize the disk,
without having it damp out all perturbations.  

No explicit central bulge was placed in the initial gas and stellar disk
components.  Thus, the model disk galaxies most closely resemble Scd
galaxies.  The presence of a central bulge can have a pronounced effect on
the large-scale gas flows that can be triggered by galaxy encounters and
collisions, as demonstrated by Mihos \&  Hernquist (1994).  They found that
a massive central bulge in a gaseous disk galaxy inhibits the streaming of
gas towards the center of a galaxy during a collision.  The disk gas is
thereby ``reserved,'' leaving it able to flow into the central regions at a
later time if the two galaxies should merge.

The set of available simulations are of collisions between a disk galaxy
and a gas-free elliptical galaxy.  In constructing these, three values of
the mass ratio between the disk galaxy and the elliptical were used, 10:1,
4:1, and 1:1. The impact parameter was varied from zero to a
length slightly larger than the radius of the disk galaxy.  In these
simulations, the initial velocity of the elliptical was oriented perpendicular to
the plane of the disk.  The initial speed of the elliptical with respect to
the disk was chosen so that the system was barely unbound (i.e., the
galaxies were in slightly hyperbolic orbits).  The initial separation
between the two galaxies was several disk radii. The simulations were
performed on a Cray-2 supercomputer at the National Center for
Supercomputing Applications (NCSA), between 1991 and 1995.

As mentioned earlier, the resolution of both the models and observations
are relevant when drawing inferences from detailed comparisons.  In the SPH
method the fluid is modeled with a set of Lagrangian markers or
``particles.'' These particles are attached to fluid elements and move
around with the gas. Properties of the gas are obtained by averaging over a
volume surrounding each point, which is characterized by a ``smoothing
length.'' Thus, the resolution of the simulated gas is limited to scales a
little larger than one smoothing length (see Balsara 1990; Monaghan 1992). 
In the models used for this study, the smoothing length is the same for all
SPH particles, and is held constant throughout the simulation. Resolution
in the calculation of the gravitational potential is limited by the size of
the grid cells used. The resolution of the cloud-in-cell interpolation
between grid cells limits the resolution to about half of a grid size. In
the simulations employed here, Gerber used a cubic volume with 64 mesh
cells on a side, where the diameter of the disk galaxy spanned about 35
cells.

The numerical computations were performed using ``computational units'' for
the dimensions of mass, length, and time.  These units can be scaled to the
physical dimensions of any real system to which the models are to be
compared, as described in Section 3.4.  The physical mass and length
resolution of a simulation scales with the physical units imposed on the
models.

A limitation of these simulations is the treatment of the gaseous material
as a one-phase medium.  With this restriction, the simulation gas can be
used to investigate only the gross flow patterns induced by a collision,
the ensuing density changes, and the occurrence of shocks.  Taking a
large-scale view of the gas motions is consistent with the limited
resolution of the modeled gas, which is typically around 750 pc for L*
galaxies.  This resolution is sufficient to allow the study of a variety of
interesting collision-related phenomena, as discussed below. 

No prescription for star formation was included in these simulations. 
Instead, the gaseous component of the disk galaxy was allowed to evolve
without mass loss or thermal feedback from newly formed stars.  For the
problem of galaxy collisions, the violent gas dynamics and long-range
gravitational forces have a much larger effect on the overall structure
than any star formation triggered in the gas, even though the star
formation can be extensive.  That is, star formation triggered by a
collision has only a second-order effect on the energetics of the system. 
However, the inclusion of a star formation algorithm appears to be critical
in codes used to study galaxy formation. These algorithms necessarily
include one or more free parameters, such as the star formation
``efficiency,'' which are needed because of the rather limited
understanding of the physics of star formation (c.f., Carraro, Lia, \& 
Chiosi 1998; Koda, Sofue, \&  Wada 2000; Thacker \&  Couchman 2000).  This
study and several others (see Gerber 1993; Gerber, Lamb, \&  Balsara 1992; 
Lamb, Hearn, \&  Gao 1998) suggest that monitoring the evolution of
the dense, gaseous structures that emerge where streams of gas collide and
shocks form (as a result of the collision) is sufficient for finding
probable regions of large-scale star formation in real colliding galaxies. 
Through the simultaneous analysis of observed star forming regions in
colliding galaxies and the locations and behavior of corresponding dense
regions in the simulated gas, we are able to learn more about the physics
of large-scale star formation.

\subsection{Computational Match to Arp~119}

Our choice of the best model to compare with Arp~119 involved the analysis of variations in the morphology of the simulated disk galaxies (projected onto the plane of the sky) due to mass ratio, impact parameter, time since collision, and viewing angle.  At first, it may appear difficult to decouple the effects of individual parameters on the final image.  However, the level of asymmetry and other details of the disk of Arp~119S galaxy give clues about the impact parameter and mass ratio of the collision.  After a head-on collision with a significant impact parameter, the nucleus of a disk galaxy often appears offset from its original position in the plane of the disk.  The amount of the shift allowed us to narrow our selection to a small range of impact parameters.  The apparent broken-ring structure indicated by the luminous arc in the northern part of Arp~119, and the diffuse nature of the material to the east, gave sufficient indication that the elliptical is about as massive as the disk galaxy.  Direct observations of the elliptical galaxy (see Section 2.1) confirm that the two members of this pair have roughly equal masses.  

The ``current epoch'' timestep in the simulation and the appropriate viewing angle were determined by matching the predominant collision-induced features of the projected simulation image with strong features observed in Arp~119S.  The most important of these features is the bright, luminous arc to the north of the nucleus that can be associated with higher density gas and newly formed stars (see Section 4).  B-band, H$\alpha$, and CO images reveal the shape and extent of this structure, as well as its position with respect to the nucleus.  Different simulation timesteps and viewing angles were surveyed until the best morphological match could be made. The viewing orientation of the model was then fine-tuned to bring a southern arc of lower-density gas into alignment with the southern arc of knots seen in Arp~119S (Figure 1).  The orientation of the disk with respect to the line-of-sight, as well as the trajectory and initial position of the elliptical, are shown in Figure 3a.  

As discussed in Section 3.1, our set of available simulations samples both mass ratio and impact parameter coursely, so no match between observations and a simulation will be exact to the resolution of the models.  However, in accordance with the philosophy discussed above for fitting computational models to observed systems, we find that a ``best fit'' simulation for Arp~119 is one of a collision between two equal mass galaxies.  (Here, the mass of each galaxy includes its dark matter halo.)  The impact parameter of the collision is about 25 percent of the disk's radius.  Figure 3d shows a representation of the present state of the gas for Arp~119, as viewed from the Earth's perspective.  (We designate this simulation timestep as the ``present time.'')  In this figure, the disk gas is displayed along with the elliptical.  The disk SPH particles are shaded according to volume mass density.  

It is important to note that the individual SPH particles represent different masses of gas.  Thus, the density of SPH particles in a particular region is not necessarily representative of the gas mass density at that location.  Instead, each particle carries its own local density information, which is used to shade the particles (see the captions for Figure 3).  The distribution of stellar disk particles for the current epoch is shown with the elliptical particles in Figure 4.  The disk halo material is not displayed in any of the figures, but is included in all of the computations.  

The system is oriented so that the disk galaxy's initial rotation axis makes a 65-degree angle with the line of sight.  This orientation agrees well with the angle of 63-degrees determined by Marziani, et al.\ (1994).  For comparison with the Earth perspective viewing angle, face-on and edge-on views of the SPH particle positions are shown in Figure 5a and Figure 5b, respectively.  The elliptical is shown at the very top of Figure 5b, but not in Figure 5a to prevent confusion in the interpretation of the gas particles.  Note the considerable spread of material out of the disk plane apparent in Figure 5b, as well as the marked amount of disk warping.  The face-on view of Figure 5a clearly demonstrates that the galaxy is a classic collisional ring galaxy.  In it we see the typical structure observed in impacted disks when a slightly off-center collision produces an off-center nucleus and a broken ring (as seen in galaxy Arp 147).

Several similarities are apparent in the structures of the model gas disk (Figure 3d) when compared to the optical image in Figure 1a.  Both have a strongly disturbed appearance, featuring an off-center nucleus.  An arc of high-density gas is present to the north of the nucleus in the model and extends to the west.  This arc is very similar to the bright arc in the northern limb of Arp~119S.  The long arc of knots at the southern edge of Arp~119S is coincident with knots forming a southern arc in the simulation.  The gas density is relatively low in the eastern half of the simulation, matching the region of low luminosity in the optical image.

As near-infrared emission from galaxies does not have major contributions from recently formed stars, either directly or via reradiation from dust, images in this wavelength region trace the distribution of the older stellar population.  That is, near-infrared observations of Arp~119 reveal the stellar population present before the collision took place.  The J- and H-band images (Figures 2a and 2b, respectively) can be usefully compared with the simulation disk stars at the present time (Figure 4).  These images indicate that the observations and model agree, in that the stellar nucleus is displaced considerably from the visual center of the disk galaxy.  Both also show a ``veil'' of low-density in the stellar distribution in the eastern part of the galaxy. 

\subsection{Physical Scaling of the Numerical Model}

The computational units used in the simulation were chosen in such a way that they could be rescaled, allowing the simulated galaxies to represent real galaxies of various masses and radii. The gravitational constant anchors the relationship between the units of mass, length, and time, with the mass and length scales taken from observation. For the Arp~119 system, the mass unit was chosen so that the total mass of the model disk galaxy is equal to the mass of Arp~119S, namely $3.7 \times 10^{11}~\msun$ (see Section 2.1).  The unit of length was set so that the furthest extent of the northwest arc in the simulation gas disk is located roughly 13 kpc from the center of the nucleus (in projection).  This distance corresponds to the 14'' radius observed by Marziani, et al.\ (1994).  This feature was chosen to set the length scale because it is well defined and has an associated, dense structure in the simulated gas disk of the ``current epoch'' model.  With this choice, the SPH smoothing length becomes 650~pc, setting the resolution scale of the simulated gas disk.  The grid spacing for the PM gravity solver becomes 910~pc, implying a resolution in the gravitational potential of approximately 450~pc.

The units for time were then calculated from the relation 
{\hbox{$K = M \times T^2  /L^3$}}, where, $M$, $T$, and $L$ are the units of mass, time, and length, respectively.  The quantity K is proportional to the gravitational constant and it's value sets the numerical scale for the simulation. The time evolution equations used produced timesteps with non-constant spacing.  However, the average space between timesteps is 2.2 Myr, with variations of only about 20\%.  Thus, the entire simulation of 70 timesteps (including 14 pre-collision timesteps) spans a scaled time of approximately 150~Myr.  

Once the physical scales of the collision simulation are set, the initial models for the disk and elliptical galaxies can be described quantitatively.  The initial gas and stellar disks have a radial scale length of 3.6~kpc, and are cut off at a radius of 16~kpc.  The scale height for the disks is 730~pc, with vertical cut-offs at about 2.9~kpc.  Because all of the elliptical galaxy's n-body particles have a uniform mass, the mass distribution of the elliptical can be described by the root-mean-square radius for the particles (see Section 3.2).  The initial elliptical model has a RMS radius of 5.8~kpc, with a radial extent of 17.4~kpc.  Similarly, the disk galaxy's halo has a RMS radius of 7.2~kpc, and a cut-off radius equal to that of the gas and stellar disks. 

With our adopted physical length unit scaling, the sizes of the projected major and minor axes of the simulated stellar disk at the present epoch are 51~kpc and 22~kpc, respectively.  These dimensions are somewhat smaller than those observed for Arp~119S, for example, the 68~kpc by 30~kpc measured by Marziani, et al.\ (1994).  This apparent discrepancy may be due to the mass of Arp~119N being somewhat larger than that of Arp~119S (see Section 2.1).  More massive ellipticals tend to increase the rate at which disk material spreads out during a collision.  The equal-mass galaxy simulation may not accurately simulate the extent to which matter is dispersed in the southeast part of the disk.  Additionally, the observations used to determine the spatial extent of Arp~119S may be sensitive to low density material that is not represented in the simulation due to the radial cut-off chosen for the gaseous and stellar disks.  This consideration indicates that one must be careful in interpreting observed outlying features of the real disk galaxy in terms of the simulations. While regions of enhanced density are modeled well by the simulation, low-density regions near the boundaries of the disk may not be.  Sensitive H~I observations, for example, may indicate the need for models with extended gas disks.  

In the simulation, the initial relative velocity between the centers-of-mass of the two galaxies is 730~km~s\sup{-1}, while the relative speed in the ``present time'' model is roughly 440~km~s\sup{-1}.  At this timestep, the centers-of-mass of the two simulated galaxies are separated by a distance of about 42~kpc.  The time since collision for the current epoch in the scaled simulation units is roughly 71~Myr.  Thus, since the collision took place, the average relative speed between the two model galaxies has been 580~km~s\sup{-1}.  

\subsection{Collision Partners of Arp~119S}

The observations suggest that Arp~119N is the most recent galaxy to have collided with Arp~119S.  Not only are the two galaxies very close to each other in the plane of the sky, but their redshifts are also quite similar.  Figure 1a shows a significant amount of luminous material in Arp~119S that appears to be drawn toward this companion.  Comparisons between the observations and the numerical model confirm this relationship between the two galaxies.  Most significantly, the location of the elliptical with respect to the disk galaxy in the current epoch model mimics the observed configuration in Arp~119.  Here, the separation between the two real galaxies is about 20\% larger than that of the model galaxies.  In addition, Arp~119N exhibits the same stretching along the direction toward Arp~119S that is seen in the elliptical during the simulated collision.  

Using the timescale information derived from the simulation together with the observed distance between the two real galaxies, the average relative speed of the galaxies in the plane of the sky over the time since the collision is approximately 730 km~s\sup{-1}. Given that the observed line-of-sight relative velocity between Arp~119S and Arp~119N may be as high as 700 km~s\sup{-1}   (Marziani, et al.\ 1994), the relative spatial velocity may be as high as 1,000 km~s\sup{-1}, averaged over the time since collision.  However, in Section 5.1, we discuss these velocities in terms of a fit to the ``present time'' model, and argue that the relative line-of-sight velocity between the two is likely closer to 500 km~s\sup{-1}. (See Section 5 for a further discussion of the velocities).  This indicates that the initial relative velocities of the two galaxies in the Arp~119 system was high, at least 850~km~s\sup{-1}, and possibly as large as 1,000~km~s\sup{-1}. Thus the real relative velocity was either slightly larger than the scaled initial relative velocity used in the simulation, or about one and a half times this value.  It is definitely reasonable for the chosen simulation to employ a hyperbolic trajectory for the elliptical with respect to the disk galaxy.  

When observations of H~I emission (C. Horellou, private communication) are taken into account, another possible collision partner becomes apparent.  Mrk 983 is a disturbed disk galaxy located about 2.6 arcminutes south of Arp~119S, a projected distance of 145 kpc.  It has a redshift equal to that of Arp~119S.  In the H~I image, a possible, faint gas bridge appears between Arp~119S and Mrk 983.  From the asymmetries in the H~I halos surrounding Mrk 983 and Arp~119S, it is evident that a glancing collision may have occurred between these two galaxies.  Indeed, Mazzarella, Bothun, \&  Boroson (1991) observed two spatially distinct H~II ``hotspots'' in Mrk 983, indicating that it has been involved in galactic interactions in the past.  

An analysis of the collision simulation provides good evidence that Arp~119N has caused the collision-induced structure seen today.  As a result of a collision, the nucleus of a disk galaxy is often shifted out of the plane of the rest of the disk in the direction of the collision partner's motion.  From a survey of different three-dimensional orientations of the simulated galaxies, it is difficult to see how the nucleus of Arp~119S could appear so close to the northern edge of the disk (in projection) while having a southward-traveling collision partner.  Furthermore, the collision timescale given by our modeling implies that an average relative speed since the collision of more than 2000 km~s\sup{-1} would be required to account for the current separation distance between Arp~119S and Mrk 983.  It is therefore possible that Arp~119S has interacted with two partners in its recent history, but if so, the observational evidence suggests that the more recent, head-on collision with Arp~119N was the more violent of the two.  The interaction of Arp~119S with Mrk 983 would have only been a glancing encounter.  The dynamics of these three galaxies suggest that they may form the heart of a compact group.  

\section{Tracing the History of Star Formation}

As discussed earlier in this paper, and in many other places in the literature, the collisional nature of the gas and the changing gravitational field during the collision produce significant, local density increases in the gaseous disk.  The combined density/material waves propagate through the gas away from the impact point, and combine with the rotation of the disk to produce well defined, identifiable structures in the gas.  Indeed, as the density waves propagate through the gas disks of the simulations, they take the form of a ring, long arc, or one or more spiral arms.  These features are similar to the luminous structures observed in collisional ring galaxies and morphologically related objects, such as the ocular galaxies, one-armed spirals, and pronounced two-armed spirals observed in some colliding pairs (see, for example, Kaufman et al.\ 1999).  Observations of colliding galaxies indicate that under some conditions, prodigious star formation can occur within these dense features (see Borne, et al.\ 1994; Whitmore \&  Schweizer 1995; Appleton \&  Marston 1997).  

The connection between the propagating gas density wave in the simulations and the occurrence of large-scale extranuclear star formation is clarified when observations taken at multiple wavelengths are compared to appropriate models.  The new stellar population created during a starburst episode will initially contain many massive, blue stars as compared to the ambient, background population, which will be older and thus redder.  Over the time of a disk galaxy collision (a few hundred million years for an L* galaxy), there is the opportunity for several waves, or generations, of starbursts to occur.  These starbursts are triggered in the highly reorganized gas by the passage of density waves and by the infall of gas towards the center of the galaxy.  The average color of the localized starburst regions will redden with time, and the spectra will become that associated with ``post-starburst'' regions (see, for example, Bernl\"{o}hr, 1993).  Starbursts are evidenced by H~II line emission and B-band stellar continuum emission dominated by O and B stars, while post-starburst emission is dominated by the continuum of late B and A stars having strong Balmer absorption and little line emission.  Thus, the variously-aged starburst populations can be discriminated using a careful comparison between observations of the galaxy in different wavelength regimes (see Gallagher, et al.\ 1984; Hunter, et al.\ 1989).  For a discussion of this as applied to colliding galaxies see Lamb, et al.\ (1998).

Our galaxy simulations give information about the location of dense gas and the occurrence of shocks as a function of time.  Thus, the current-epoch simulation gas disk of Figure 3d can be used to show where the current locations of dense gas are to be expected in the real galaxy.  Observations of the molecular gas in Arp~119S (Gao 1996; see Section 4.4) confirm these expectations.  Similarly, studies at other wavelengths can reveal the prior regions of dense gas and shocks by locating the remains of starbursts triggered at these earlier times.  The new stellar populations that formed progressively as a result of the galaxy collision are not expected to be fully coincident with the current high density regions in the gas.  Rather, these new stars mark the places where the dense gas was located (relative to the underlying rotating galaxy) at the time they were formed, because the density wave leaves the material behind.  The starburst regions will have participated in the overall rotation of the galaxy since their formation.  However, they will have moved, at most, through about 20\% of their orbit by the time we observe them because of the slowed rotation rate in the disk behind the outwardly expanding density wave (see Lamb et al.\, 2000).  Thus, because of the slight uncertainties in the locations of formation for observed post-starburst regions, as well as the uncertainty in the position of dense regions in the simulated gas (approximately 650~pc), there is an uncertainty in the age of the starburst, as determined from the simulation.  We estimate this uncertainty to be on the order of ±3 Myr.  The more recent the starburst, the more accurately we can date it, as is the case when purely-observational methods are used (see Gallagher et al.\, 1984).

In summary, observations at various wavelengths, used together with simulations, allow us to run the real galaxy collision ``backwards'' to see where and when previous star formation episodes occurred.  That is, we can compare the locations of the young, massive stars with those of shocked, dense, gaseous structures appearing in earlier timesteps of the simulation.  We can then gain some understanding of the timescales involved in these collision-induced starburst events, and of the physical conditions in the gas when star formation occurred.

\subsection{Important Features of Arp~119S}

The most pronounced feature in Arp~119S at all wavelengths, including the B-band, is its off-center nucleus.  Its presence in the B-band and H$\alpha$ observations indicates that the nucleus has experienced extensive star formation since the collision took place. The first significant effect of the impact of an intruding galaxy on a disk galaxy, especially in an almost head-on, high speed encounter like this one, is to rapidly pull the stars and gas towards the center-of-gravity of the combined system.  As the intruder moves away, the non-collisional stars do not remain in this contracted state, but, conserving angular momentum, flow out to even larger radii than they had before the encounter.  However, the collisional nature of the gas ensures that some of the gaseous material will remain in the nuclear region, presumably to form stars under some circumstances.  Further, as asymmetric structures form in the disk, there is a tendency for gas to flow inwards towards the center.  

A second dominant feature observed in this galaxy is the bright knot just to the north of the nucleus.  This structure is prominant in images at the four different (non-infrared) wavelength bands currently available to us, and may also be visible in the near-infrared,  causing the elongated contours surrounding the nucleus in our J-band image (Figure 2a).  Both the location and the velocity structure of this knot are taken into account when interpreting it in terms of the galaxy collision simulations, as we describe later in this paper. 

A ``cusp'' of emission is observed at the northern-most extent of Arp~119S in the visible, B-band, and H$\alpha$ images, but not in the CO image.  This is clearly a region of moderately recent star formation (because of the presence of some H$\alpha$ emission).  A broader, but still well-defined structure is also visible at this same location in the near infrared data.  This formation has no obvious counterpart in our current simulation.  However, the very high relative velocity obtained for this structure by Marziani, et al.\ (1994), as well as its alignment with the nuclei of the two galaxies, point toward its connection with the elliptical galaxy.  We discuss the likely source of this structure in the context of the dynamics of the system in Section 5.  

In the remainder of this section, we examine the B-band, H$\alpha$, and CO images of Arp~119S in comparison with our simulation.  From this analysis we can draw some conclusions about the star formation history and the evolving structure of the disk galaxy.

\subsection{B-band Observations: Starburst and Post-Starburst Regions}

The new stellar population created during a starburst episode will contain a much larger number of massive, blue stars than the original (older) population.  Figure 1b shows a B-band image of an interior region of Arp~119S obtained by Marziani, et al.\ (1994), which highlights the locations of the stars formed since the collision took place.  Both very recent star formation regions and those in a post-starburst phase contribute to this band.  Thus, the B-band flux provides a measure of the amount of star formation integrated over the last few tens of millions of years (see Gallagher \&  Hunter 1987).  

One important feature of this image is the distinct, complete ring encircling the nucleus.  The ring has a radius of about 6'' (Marziani, et al.\ 1994), or 5.6~kpc, and appears to be coincident with the one seen in the optical image (Figure 1a).  In our ``best fit'' simulation, a similar structure appears in the disk gas soon after the collision takes place.  At a time 24~Myr after the collision, the ring in the simulation has reached a radius roughly equal to that of the observed 6'' ring.  The gas disk from this simulation timestep is shown in Figure 3b.  The collision between the elliptical and the disk galaxy produced a density wave in the gas that initially propagated outward in this ring shape.  Thus, a comparison between the B-band image and our simulation strongly suggests that there was an intense burst of star formation in a dense ring structure approximately 24~Myr after the collision (that is, 47~Myr before the present time).  This ring is now entering a post-starburst phase while other parts of the galaxy are currently experiencing prodigious, large-scale star formation.  

The B-band observations prominently display a broad arc to the north of the nucleus, extending to the west.  This structure is fit well by the simulation's gas disk in the current epoch (see Figure 3d).  This fit suggests that this region contains a significant portion of the very youngest stars in Arp~119S, which is confirmed by the H$\alpha$ image.  

Discrete regions of emission are seen in the southern and southwestern parts of Figure 1b.  These regions appear to be coincident with knots seen along the southern edge of Arp~119S in the optical image (Figure 1a).  Thus, there is evidence that the southern line of knots have significant levels of B-band emission, and may be newly-formed super star clusters.  These clusters come within about 16'' (15~kpc) of the nucleus, in projection, and they are the most likely candidates for proto-globular clusters in this galaxy.  Figure 3c shows the simulation's gas disk at a time 46~Myr after the collision, about 24~Myr before the current epoch.  Here, a set of enhanced-density knots is visible to the south and southwest of the nucleus.  These knots are apparent in the simulations over a considerable time, so the temporal origin of the observed knots is not closely constrained by the simulation.  We have chosen this timestep (Figure 3c) to illustrate that the current simulation captures behavior that may well be associated with the formation of globular clusters in real galaxies.

\subsection{H$\alpha$ Observations: Present Day Starburst Regions}

H$\alpha$ emission is associated with the ionization of the interstellar medium in regions of strong ultraviolet flux.  Outside of the nucleus, this emission is always found in the vicinity of young, hot, massive stars.  Thus, the large regions of H$\alpha$ emission found in Arp~119S by Marziani, et al.\ (1994), and shown in their H$\alpha$ image (see Figure 1c), can be used to find the locations of the most recent bursts of star formation in Arp~119S.  In this image, there is a northern arc in the same general region as the one seen in the blue.  However, the H$\alpha$-arc appears to have a somewhat stronger signal to the east, and a slightly weaker one to the west, compared to the luminosity distribution of the B-band image.  This shift in position is explained by noting that the strongest peaks in the simulation's gas density wave migrate around the expanding arc, from the western side of the nucleus where the impact occurred, towards the east.  Thus we expect that the peak in the star formation rate along the upper, bright rim of the disk would also have migrated from the west towards the east during the last 50 million years.  The H$\alpha$-arc is modeled well by the distribution of the densest gas in the ``present day'' simulation model (Figure 3d).  (Since the temporal resolution of the simulation is about 2 Myr, the current epoch timestep should provide the closest match to both the molecular gas and H$\alpha$ observations.)  

The moderately intense region southeast of the nucleus in the H$\alpha$ image has no clear counterpart in the B-band image.  However, a similar structure is seen in the CO image (see Section 4.4).  The strength of the H$\alpha$ signal indicates that this is another region of recent star formation, but no such structure appears in the simulation gas.  This observed region of dense gas appears connected to the nucleus, and it may result from the passage of the elliptical through a more extended, dense gaseous disk than was employed in the simulation.  The lack of a strong corresponding signal in the visual and B-band images indicates that this starburst is very young, of the order of only a few million years.

We note that the prominent 6'' ring of the B-band image is not present in the H$\alpha$ image, and that there is no clear evidence for the southern arc of knots, either.  This absence is consistent with the idea that the stars in these features formed at an earlier time.  

\subsection{CO Observations: Dense Gas at the Current Epoch}

Before new stars can be produced, a region of cool, dense gas amenable to collapse must form.  An indication that such an environment exists is the presence of a significant amount of molecular hydrogen.  Unfortunately, H\sub{2} has essentially no emission in the cool (10~K) environments of molecular clouds.  However, \sup{12}CO(1--0) transition can act as a suitable tracer for cool regions of H\sub{2} (see also Solomon, et al.\ 1997; Young 1997).  There is debate about the correct factor to use in converting CO fluxes to H\sub{2} masses, and whether this factor is the same in all circumstances.  However, the presence of CO does indicate that the cool, dense conditions necessary in the gas for star formation exist at a specific location in a galaxy.

We use the CO (1--0) observations of Gao (1996) to reveal the regions in Arp~119S where further large-scale star formation may occur (see Figure 1d). The eastern part of the northern arc as seen in the B-band and H$\alpha$ images, is even more pronounced in the CO image.  We interpret the significant eastern CO extension as indicating that the strongest density peak in the gas along the northern rim of the disk has now migrated through over a quarter of the ring circumference since it was formed by the impact.

There is also an extended CO emission region southeast of the nucleus.  This arc of gas appears to stretch about half of the distance from the nucleus to the region of the southern arc of knots as seen in Figures 1a and 1b.  The eastern extent of this southern CO arc curves southward, and coincides with the strongest part of the H$\alpha$ emission in this region (see Figure 1c).

\section{The Velocity Structure}

\subsection{Observed Disk Motion compared to Model Predictions}

In a quiescent disk galaxy, a typical element of the disk material can be thought of as executing epicyclic motion about a nearly circular galactic orbit.  There is usually little or no bulk flow of material in the radial direction.  Thus, for an appropriately oriented galaxy, one can determine azimuthal (orbital) velocities for the disk using line-of-sight velocities along the galaxy's major axis. 

A collisional disk galaxy may exhibit some significantly different kinematics.  In the type of collision studied here, that is, an almost head-on collision parallel to the disk spin axis, radial velocities in the disk can be as large or larger than the azimuthal velocities.  Motion towards and away from the center-of-mass occurs.  After the passage of the density wave, the azimuthal velocity of an element of disk material is lower than that of material previously located at the same radius.  The azimuthal velocity at the peak of the density wave is a local maxima, and often the global maximum (Gerber 1993; Gerber, et al.\ 1996; Lamb et al.\ 2000).  

We note that while the simulation particle speeds should closely resemble the measured velocities in a real galaxy, they do not indicate the speed of the collision-induced density wave.  The pattern speed is larger than the material speed in many instances, so care must be taken when trying to deduce the wave propagation speed from the observations.  

The physical scaling of the numerical model, described in Section 3.4, does not involve the observed velocity data, nor were velocities used in selecting the simulation fit to the system.  Thus, an analysis of the velocities in the current epoch model provides further information that can be compared to the observed velocities, and used to study the dynamical history.  The graphs shown in Figure 6 provide two-dimensional velocity information for the simulated disk gas at the current epoch.  In these graphs, the disk is viewed face-on from the side of approach of the elliptical.  The elliptical's center-of-mass passed through the disk to the right of the disk center, along the horizontal axis.  The disk material is rotating counter-clockwise in this view, as it is in the Earth-perspective view (Figure 3).  To aid in interpreting these velocity contours, we provide a dashed line across each plot to indicate the position of the major axis of the simulation disk as viewed from the Earth perspective.  That is, although these graphs provide a face-on view of the galaxy, the orientation is not very different from the Earth view of Figure 3, and the graphs can be used to visualize the components of velocity in the disk along our line-of-sight.

Figure 6a shows the in-plane azimuthal velocities across the gas disk surface, where counter-clockwise motion is positive. Figure 6b displays the in-plane radial velocities across the surface of the gas disk, where outward motion is positive.  In both graphs, the radial and azimuthal components are determined using the disk galaxy's center of mass (including stellar, gas, and halo material) as the coordinate origin, and only those velocity components parallel to the pre-collision gas disk plane are used.  The graphs were constructed by decomposing the plane of the disk into a 2-dimensional array of pixels.  For each pixel, the reported velocity is the mass-weighted average of the velocities for all SPH particles in the column normal to the disk plane.  (The cross-section of each column, or pixel, is a little larger than two smoothing lengths across.)  Figure 6c shows the ratio of the radial velocity component to the azimuthal component at each pixel along the disk surface, and can be used to determine the dominant component of velocity at that position in the disk.  

The azimuthal velocity graph (Figure 6a) shows the asymmetries in the azimuthal velocities that resulted from an off-center collision.  To the right of the nucleus, the rotational velocities along the major axis are in the 200--300~km~s\sup{-1} range, and the peak in the azimuthal velocity of over 300~km~s\sup{-1} occurs in a small region to the right (west) of the nucleus and below (south) of the major axis.  The overall azimuthal velocity structure reflects the result of the continued inspiraling of the outer disk material towards the impact point, as it collides with outward-spiraling material just inside the ring.  The high density features are produced where the two flows meet. The radial velocity graph (Figure 6b) displays a wide range of expansion and contraction velocities. In the current epoch, material near the nuclear region is generally contracting with velocities as high as 100~km~s\sup{-1}.  Material near the low density ``veil'' (towards the left side of this figure) is expanding as fast as 400~km~s\sup{-1}. 

The ratio of the magnitude of the radial velocity component to that of the azimuthal component, across the disk surface, is illustrated in Figure 6c.  Throughout much of the central region of the disk, the radial velocities are less than the azimuthal velocities, even though the latter are depressed compared to what they were at the same locations in the disk before the collision.  Towards the edge of the disk, the radial velocities can be equal to, or greater than, the azimuthal velocities.  Along the right side of Figure 6c (western edge of the disk), the radial velocities are up to twice as much as the azimuthal velocities, while along the left side (eastern edge) there is an even higher ratio of radial-to-azimuthal velocities.

In Figure 7 we show a line-of-sight velocity field for the simulated gas disk.  Here, the disk is viewed from the Earth-perspective at the current epoch.  These contours were generated from a dataset composed of a two-dimensional array of pixels.  Each pixel has a width equal to the smoothing length of the SPH particles, and represents a column of gas along the line of sight. The velocity associated with each pixel is the mass-weighted average of the gas particles for the column, and overlapping gas populations with distinct velocities are averaged together.  The contours display a continuous gradient in the line-of-sight velocities across the disk galaxy.  Both the azimuthal and radial velocity components are significant in this velocity field.  The line-of-sight velocities of this graph include contributions from the full velocity of each SPH particle, not only those components in the plane of the disk, as in Figure 6.  Thus, the velocities in Figures 6 and 7 are not directly comparable.  Because the H$\alpha$ flux originates in stars that have formed in the gas, not the gas itself, and the star formation rate is not a smooth function of time in this system (see Section 6.1), it may not be appropriate to make a direct comparison between the observed line-of-sight H$\alpha$ velocities and those constructed from the model, as displayed in Figure 7. Rather, we interpret the observed velocities near the rim of the disk only in terms of the in-plane velocities of the simulated gas disk, as displayed in Figure 6.

Below we present a detailed comparison of the observed velocities with those found from the model as viewed at an inclination angle of 65 degrees (see Figure 3). Vital to the interpretation of the observational data is the adopted value of the systemic velocity of the Arp~119S galaxy. From a detailed comparison of all the available velocity data, we find that a fit is possible only if the systemic velocity is approximately 14,440 km~s\sup{-1}. This value is to be compared with the values of 14,200~km~s\sup{-1}, as measured by Marziani, et al.\ (1994) using sodium D lines, and 14,380 km~s\sup{-1} from Mazzarella \& Boroson (1993).  Therefore, we have adopted a value of 14,400 km~s\sup{-1} as the systemic velocity throughout the remainder of this discussion.

There are two long-slit spectra providing velocity data relevant to the disk velocities (see Marziani et al.\ 1994). In one case the slit was placed along the major axis, as viewed from Earth, and in the other, it was placed along the northern rim, roughly tangentially. The resulting velocities are displayed in their Figures 5 and 6.  Making a correspondence between the model and Arp~119S, the arc of dense gas along the northwestern limb of the disk galaxy is equivalent to a location above, and to the right of the nucleus in this view of the model, about three-quarters of the way to the edge of the disk. 

In the model, the location corresponding to the dense northwestern arc has radial gas velocities around 200~km~s\sup{-1}, directed outward, comparable in magnitude to the azimuthal velocities in that region.  Given, the high inclination angle of Arp~119S (about 65-degrees), our line-of-sight velocity across this rim should be roughly constant, with the material at the most western end contributing a component of its azimuthal velocity towards us, and the material closest to the nucleus contributing a comparably sized component derived from the radial outflow in the disk.  That is, we predict a roughly flat rotation curve across this northern rim of the disk, as displayed in Marziani, et al.\ (1994) Figure 6. The eastern edge of the detected emission corresponds to a position in the model disk where the azimuthal velocity has a component away from us, and this shows up in the observational data as a slight rise in the recessional velocity.  If the systemic velocity of Arp~119S is 14,400~km~s\sup{-1}, the velocities observed in this arc (roughly 14,280~km~s\sup{-1}) are consistent wtih the 100--200~km~s\sup{-1} expansion velocities seen in this region of the model.

The velocity structure along the major axis shows considerably more variation than that found along the northern rim (Figure 5 of Marziani, et al.\ 1994).  Here, we limit the discussion to the material which is found across the nucleus and out to the western rim, beyond the bright, northwestern arc (labeled region ``R'' in the observational data), where velocities are in the approximate range of 14,100 km~s\sup{-1} to 14,350 km~s\sup{-1}.  We associate these velocities with material flow within the disk. Near the place where the major axis of the disk galaxy crosses the northwest arc, the velocity curve is very flat, with heliocentric values of about 14,300 km~s\sup{-1}.  Again, we interpret this flat portion of the rotation curve as due to the combination of the components of radial outflow and rotation around the ring.

A high radial expansion rate in material flowing towards the Earth may have caused the ``dip'' in the velocity curve near the nucleus in this spectrum.  We do not interpret this as due to motions in the disk plane because the flow in the inner disk is directed towards the nucleus at this time in the simulation.  Rather, we interpret this as due to a contribution from material flowing radially outwards from the nucleus, in directions out of the disk plane (see Lamb, Hearn, \&  Gerber, 2000). The magnitudes of these velocities are similar to the radial in-plane velocities seen in Figure 6b. If we are observing flux from material out at a distance from the nucleus comparable to the disk radius, the velocities will be near the 14,200 km~s\sup{-1} value at the minimum of the dip.  

The major axis slit position of Marziani, et al.\ (1994) passes through the nuclear region and includes a bright knot just to the east of the nucleus (see the B-band image of Figure 1b).  This knot is coincident, in the plane of the sky, with the western extent of the 6''--7'' ring. Two velocity components are found in this combined nuclear region, one around 14,500~km~s\sup{-1}, corresponding to a relative velocity away from us of 100~km~s\sup{-1} in the galaxy frame of reference, and one at about 14,200~km~s\sup{-1}, corresponding to a velocity of 200~km~s\sup{-1} towards us.  No signal is detected to the east of the bright knot, but to the west the velocity gradually decreases from the 100 km~s\sup{-1} towards us, to approximately the systemic velocity.  Again, this range of velocities can be interpreted as projections of the radial, out-of-plane flow of disk material onto our line-of-sight.  The gradually changing velocity to the west of the nucleus reflects the changing projection of the radial outflow in the disk.  The bright knot to the east of the nucleus has no direct analogy in our models (see Figures 1b and 3b).  We suspect that it is due to one of the strongest bursts of star formation that occurred in the expanding ring-like density wave.  

\subsection{Peculiar Velocity Systems: Evidence for a Colliding Gaseous Elliptical?}

The similarities between the observations of Arp~119S and the numerical simulation indicate the robustness of our chosen model.  The locations of large-scale star formation correlate with the propagating collision-induced density wave in the model disk, and the observed velocity structure along the rim of the northern disk is consistent with the velocity structure in the model.  However, there are differences between the model and the observations that may allow a deeper understanding of some special aspects of the collision that occurred in the Arp~119 system.  A major difference concerns the observed high velocity gas, found at locations near the minor axis of the disk, in and to the north of the nuclear region. 

Specifically, Marziani, et al.\ (1994) observed various distinct velocity systems in H$\alpha$, along the minor axis of Arp~119S (see Section 2.1).  A spectrograph slit placed along this direction passes through both the nucleus of Arp~119S and through Arp~119N (see Figures 1 and 4 in their paper).  The observed velocities range from approximately 14,100 to 15,500~km~s\sup{-1}, which bracket the likely systemic velocity of the disk galaxy of 14,400~km~s\sup{-1} that we deduced by combining observed velocities in the northern rim with results from the numerical model (see Section 5.1).  Thus, we estimate that the various velocity components have line-of-sight velocities ranging from -300 to 1100~km~s\sup{-1}  , with respect to the center of mass of Arp~119S.  In their observations of this object, Osterbrock \&  Dahari (1983) observed two different emission line profiles, each with its own set of relative line strengths.  These authors felt that the spectra provided evidence for two nuclei in the disk galaxy, with a relative velocity difference of 840~km~s\sup{-1}.  Our modeling does not support this latter hypothesis. 

The large range of velocities observed in H$\alpha$ along the minor axis of Arp~119S, can be attributed to radial outflow of shocked gas and newly formed stars.  The radial in-plane motion of material in the disk will contribute to the velocities along the line-of-sight.  However, since the radial outflow has a spherical component that includes a velocity component normal to the disk plane, the high density of material in the nuclear region ensures that there will be considerable outflow, both above and below the plane of the disk galaxy.  The simulations show rapid expansion of all components of an impacted galaxy as the other galaxy moves away.  For our chosen simulation at the current epoch, viewed at an inclination angle of 65-degrees, the flow due to the radial expansion of material along the minor axis has components directed away from Earth that range from about -200 to 300~km~s\sup{-1}. This range can be compared with that of approximately 200~km~s\sup{-1} to 0~km~s\sup{-1} (our adopted systemic velocity), measured by Marziani et al.\ (1994), for the low velocity component across the nuclear region.  Thus, the bright knot between the nucleus and the cusp of Arp~119S has a line-of-sight velocity roughly equal to the systemic velocity of the galaxy. This is consistent with the current epoch model, which shows radial in-plane expansion rates between about
-100 and 0~km~s\sup{-1} for material near the knot position (see figure 6b).  Our simulation contains no counterpart to the high velocity material. That is, although our model does predict material streaming away from the nucleus of the disk galaxy towards the elliptical, the observed high velocity system in the disk of Arp~119S reaches velocities significantly higher than those found in the simulation.  

Most likely, the high velocity material is due to the shocking of disk gas by gas stripped from the elliptical. Arp~119N is particularly devoid of gas, as evidenced by a map of the H I of this system (C. Horellou, private communication).  The highest velocities would be produced in the first gas encountered in the disk, that is, on the underside of the disk (towards the south).  As gas was stripped from the elliptical, and mutual gravity slowed the elliptical with respect to the disk galaxy, the gas shocked subsequently would have a slightly lower velocity.  Such an effect appears to be present in the Marziani, et al.\ (1994) data for the highest velocity system in the disk of Arp~119S.  In this system, the H$\alpha$ emission from the southern parts of the galaxy has, on average, a higher velocity along the line of sight than that from the northern parts, by about 200~km~s\sup{-1}. The high densities and shocking in the gas appears to have triggered prolific amounts of continuing star formation along the trajectory of the elliptical through the gas disk, as shown by the strength of the H$\alpha$ emission along the north-south (minor) axis of Arp~119S.  

The interpretation of the high-velocity system in Arp~119S as resulting, in part, from gas stripped from the elliptical is further bolstered by consideration of the northern cusp of Arp~119S (see Section 2.2; Figure 1a).  This unusual structure lies between the nuclei of Arp~119S and Arp~119N, and appears to be located along the trajectory of the elliptical galaxy.  It is apparent in the optical and B-band images, but is barely detected in H$\alpha$.  This formation has no counterpart in our chosen simulation, nor in any of those available to us, which include only a gas-free elliptical.  The H$\alpha$ spectra show a faint signal for the high-velocity system in the region of this cusp, but do not offer any clear evidence for the existence of the systemic velocity component.  

To explain their observations of the cusp material, Marziani, et al.\ (1994) argue that some of the gas in Arp~119S must have been stripped during the collision, and accelerated towards the elliptical.  They suggest that the presence of this stripped, shocked gas would account for the H~II-like line ratios seen in the northern cusp.  We add to this idea by suggesting that the cusp material consists of a combination of shocked gas from both the elliptical and the disk, which experienced a starburst shortly after the transit of the elliptical.  After the encounter, the gas and newly formed stars would have had considerable momentum in the direction of the elliptical, which can explain the high velocity observed now.  That is, as suggested by Marziani, et al.\ (1994), the orbital energy of the two galaxies could have been transferred into the interstellar medium, producing the high velocity differences seen in the disk of Arp~119S and in the cusp. 

The near-IR image of Bushouse \&  Stanford (1992), and the new ones we presented in Section 2.3, indicate that the old stellar population also has a structure very similar to the cusp seen in the B-band and the optical.  This structure is coincident with the cusp, but has a much broader, less defined profile.  It is important to note, however, that this near-infrared formation can not be the result of gas stripping.  We suggest that it is a projection effect due to a highly warped stellar disk. 

\section{Discussion and Conclusions}

\subsection{Episodic Star Formation}

In this paper, we have presented a 3D, n-body/hydrodynamical numerical simulation that closely approximates the galaxy collision that occurred in Arp~119.  It models the collision of an elliptical with a disk galaxy, where the collision is almost head-on, and is parallel to the disk spin axis.  By scaling the simulation results to the observed size and mass of the disk galaxy Arp~119S, a collision timescale was calculated.  This timescale indicates that the collision between Arp~119N and Arp~119S took place about 71 million years ago.  The simulation model that corresponds to the current epoch exhibits many of the features seen in observations of Arp~119S, both in the stars and in the gas.  This peculiar galaxy has an off-center nucleus, a bright arc of gas to the north, a line of knots to the south, and diffuse, luminous material filling the eastern half of the galaxy.

Observational images taken at a variety of wavelengths reveal the locations of past and present starburst events in Arp~119S.  We have correlated these starburst regions in space and time with regions of enhanced density and strong shocks in the simulated gas disk.  We have shown that at least three distinct star formation episodes have occurred during the collision process, one about 24~Myr after the collision, one at roughly 46~Myr, and one that is still proceeding in the current epoch.  The first occurred while the expanding ring of dense gas had not yet traveled far past the nucleus, producing the bright ring seen in the B-band image.  The second happened as the density wave neared the southern edge of the galaxy, producing the southern arc of knots.  The density pattern in the gas of the current epoch is still pronounced, and is forming large numbers of stars at the northern edge, as evidenced by the northern arc of molecular gas leading the bright arc of B-band and H$\alpha$ emission.  Thus, the evolving gaseous arc structure in the simulation provides a means of modeling the progression of collision-induced star formation in disk galaxies. 

A number of collision simulations have shown that multiple, spatially distinct ring-like waves can result from a single collision between two galaxies (for example, see Hernquist \&  Weil 1993; Lamb, et al.\ 2000).  It has been suggested that the blue 6'' ring, the northern arc, and the southern line of knots may indicate that two or three such waves are present in the disk of Arp~119S (see Marziani et al.\ 1994).  However, while various simulations by different authors indicate that two waves can form in the old stellar population, as is observed in the Cartwheel galaxy (Borne, et al.\ 1994), only one density wave forms in the gas (Gerber 1993).  It has been found that the parameter range that produces two circular density waves in the disk stars is small, in that the mass of the disk galaxy must be around three or four times that of the intruding galaxy.  In the case of Arp~119, the two galaxies have almost equal masses, and the ring and associated structures clearly represent the past history of the gas density, rather than that of the old stars.

We suggest that the various star formation regions that vary in age, and form the morphological structures in Arp~119S, are simply due to distinct star formation episodes occurring within a single expanding, ring-like wave.  That the single density wave that propagates through the gaseous disk quickly takes the form of an expanding ring, but later evolves into the more complex structure illustrated in Figures 3 and 5.  By comparing the observations with the simulation, we infer that the time interval between large-scale star formation events was approximately 25 -- 30~Myr, and that the individual bursts lasted about 10 -- 20~Myr.  The locations of the observed star formation regions are distinct, even though the burst duration is almost as long as the interval between bursts, because the density wave moves outward at a higher speed than the material.  The outward velocity of the density wave is correlated with the mass of the intruder (see Gerber 1993), where a high mass results in high material velocities and propagation speed for the density wave.  Thus, the distinct rings or regions of star formation are much more likely to be found in a colliding system in which the galaxies have roughly comparable mass, such as Arp~119, than one in which the intruding galaxy is of significantly lower mass than the disk galaxy. 

The timescale for the formation of the starbursts agrees well with that of Bernl\"{o}hr (1993) who found typical timescales to be on the order of 20 Myr in a sample of galaxies.  The discrete star formation episodes indicate that factors beyond the large-scale gas densities are important in determining star formation rates.  This implies that simple prescriptions for the star formation rate as a function of volume gas density, as employed in some simulations, might seriously overestimate the star formation rate in some galactic locations, and underestimate it in others.  Our simulation suggests that shocks as fast as Mach 10 may be present in the gas (Gerber 1993; Gerber et al.\ 1996).  It is likely that there are shock strength regimes that are conducive to star formation, while others are disruptive.  Could feedback from star formation episodes themselves cause the cessation of star formation until the density peak has moved several kiloparsecs from the location of the burst?  

There are very strong indications that episodic star formation also occurred in the disk of NGC 1144 after its collision with the elliptical galaxy NGC 1143.  The collision-induced star formation first appeared at the impact point in the disk, and followed the high density features in the gas after the collision (see Lamb et al.\, 1998).  Following the apparent correspondence of strong radio continuum emission from an off-nucleus region in NGC 1144 with the disk impact point, we predict that radio continuum observations of Arp~119S would also show high intensity from supernova remnants, in both the nucleus and the high density impact region. If found, the extranuclear supernovae burst would be indicative of an intense, prompt starburst at the location of disk impact within the first few million years after the collision.

Both the observations and the model indicate locations where proto-globular clusters may form.  For an extended period during the evolution of the density wave, clumping along the dense structures occurs on scales of a kiloparsec.  The clumps in the simulation correspond to the observed locations of distinct post-starburst regions in Arp~119S, which appear as regions of high intensity in the B-band and visible light images.  If these clumps mark the typical locations for globular cluster formation in colliding galaxies, we expect that the velocity components perpendicular to the disk plane for the star clusters will be less than the components in the plane.  However, the general asymmetric warping of the disk would certainly provide some velocity component parallel to the disk spin axis.  The closer the clumps form to the nucleus, the more isotropic are their velocity vectors, even though they form in the original disk material.

\subsection{System Dynamics}

If the Na~D-line measurements of Marziani et al.\ (1994) indicte too small a velocity for Arp~119S, then the 700~km~s\sup{-1} velocity difference between Arp~119S and Arp~119N may be too large by as much as a few hundred km~s\sup{-1}.  Still, the remaining line-of-sight velocity and the inferred velocity of Arp~119N in the plane of the sky imply a high-speed encounter of at least 700~km~s\sup{-1}, and probably close to~800 km~s\sup{-1}.  The view of Arp~119S from Earth is of the underside of the impacted disk, with the elliptical having passed up and through the disk slightly to the west of the nucleus.  The interaction would have reduced the relative velocity between the galaxies through dynamical friction, so the observed velocity difference in the line-of-sight direction of at least 170~km~s\sup{-1} (Mazzarella \& Boroson 1993) is a lower bound on the initial relative velocity in this direction. 

The detailed velocity information obtained in H$\alpha$ by Marziani, et al.\ (1994), using three different long-slit positions across the Arp~119 system, provides considerable constraint on the model fit.  The observed dynamics of the disk rim can be fit well by a disk of gas and stars that is expanding radially outward as fast as it is rotating.  The streaming motions of several hundred km~s\sup{-1} seen along the minor axis of Arp~119S can be explained as the result of the outward, roughly spherical flow produced in the central region of the galaxy that is present in the gas and the stars.  However, the simulation does not predict the observed asymmetric outflow on the order of 1000~km~s\sup{-1} from the central region of the disk galaxy in the direction of the elliptical.  Consideration of this high velocity material leads us to suggest that the presently gas-free elliptical, Arp~119N, originally had a considerable gaseous component that was stripped during the collision, and now contributes to the high velocity component observed in Arp~119S.  This theory builds on the observation of Marziani et al.\ (1994) that the line ratios determined for the high velocity cusp material indicate the presence of shocked gas, as well as their suggestion that this material was stripped from the disk galaxy. 

In this high-speed interaction, the elliptical's gas would have collided with that of the disk galaxy.  The collisional, hydrodynamic nature of the gas would have produced sharp, rapid changes in velocities and densities.  These effects would have been much more pronounced than any in the collisionless stellar components.  Consequently, the elliptical might have been stripped of its gas, while the gas in the disk would have been drawn in the direction of the elliptical's motion.  Because of the collision geometry (see Figure 3a), the affected gas would be found at a higher line-of-sight velocity than the rest of the disk galaxy. The gas would have experienced strong shocks, and possibly large-scale star formation.    

This shocked, stripped gas scenario is consistent with the prominent H$\alpha$ emission seen in the highest velocity system.  The northern cusp seen in the B-band and visual images could mark the furthest extent of this shocked gas, drawn far enough toward the elliptical to be seen beyond the edge of the disk (as projected onto the sky).  The CO(1--0) and H$\alpha$ emission region to the southeast of the nucleus may result from the passage of the elliptical through a thick, dense gas disk.  The motion of the elliptical though the disk may have produced a series of star formation bursts and shocked gas along its trajectory.  If the elliptical originally had significant gas, this component would have experienced a continuous deceleration as it passed through the gas disk.  The long slit spectra of Marziani, et al.\ (1994) suggest that the high velocity component has a decreasing velocity, relative to the disk's center-of-mass, along the direction of the elliptical's path through Arp~119S.

\subsection{Future Work}

An investigation of the role of collisions between gaseous components during galaxy interactions is central to understanding many major and novel phenomena.  The subject of large-scale star formation is addressed in this paper, and the type of study presented here can be expanded to investigations into the production of dwarf satellite galaxies in certain types of galaxy collisions and to the production of hot halos in recent merged galaxies, for example.

The gaseous component of a disk galaxy can extend to much greater radii than the stellar component (see Bosma 1981).  Recent observations (Y. Gao, private communication) suggest that strong interactions between gas disks in spiral-spiral collisions can occur well before the visible disks coincide.  This phenomenon requires that the gas disk extend significantly farther than its optical counterpart, unlike the gas disk used in the present collision simulation.  Combes (1997) suggests that the very long tidal tails found in some colliding galaxy systems are due to extensive, pre-existing gaseous disks.  We expect that the use of larger gaseous disks in collision simulations would lead to a very pronounced warping in the disk, especially for trajectories that are not parallel to the spin axis, and to the possibility of significant isolated high density regions, well beyond the old stellar disk, that might correspond to apparently-extragalactic starburst regions. 

Currently-available and planned observing facilities will allow very detailed investigations of interesting systems.  For example, future X-ray studies with the Chandra X-ray Observatory of systems that may have experienced gas-gas collisions will provide detailed information about their hot components.  These studies will help us understand the processes that produce the hot halos observed in some galaxies, such as Arp 220 (see Heckman, et al.\ 1996).  Detailed collision simulations that include a treatment of the hot interstellar material, together with an adequate representation of strong shocks, will be important for interpreting these results.

Because of a lack of observational resolution in the near-IR, the detailed dynamics of the old, underlying stellar population cannot yet be used to constrain models for many systems.  Ground-based, laser guide star adaptive optics facilities will be used to obtain much higher resolution near-IR images, and thus provide a much sharper picture of the effect of a collision on the massive stellar component.

For comparison with these higher resolution observations, new, higher resolution simulations will be performed that will allow us to interpret many observed phenomena in considerably more detail.  For example, we will be able to study the star-forming regions at smaller scales, and investigate the formation and evolution of possible proto-globular clusters more precisely.  The resolution of the hydrodynamics and gravitational potential can be increased several fold as computational storage space increases and parallel computing is fully utilized.

\acknowledgments

We wish to thank J.\ Sulentic, D.\ Dultzin-Hacyan, and P.\ Marziani for discussions on the Arp~119 system, and Yu Gao, R.\ Gruendl, and H.\ Toledo for helpful discussions concerning observations of colliding galaxy systems.  C.\ Horellou provided very thorough and helpful feedback on this paper.  We are indebted to Robert Gruendl for extensive help with the IR observations at Mt. Laguna observatory, and with the subsequent reduction of data, and to Yu Gao and C.\ Horellou who shared observations in advance of publication.
      
Part of this work was performed while SAL was a visitor at the Aspen Center for Physics.  The authors wish to acknowledge the financial support of DOE grant LLNL B506657, and of the University of Illinois Research Board.  This research has made use of NASA's Astrophysics Data System Bibliographic Services.  Use has also been made of the NASA/IPAC Extragalactic Database (NED), which is operated by the Jet Propulsion Laboratory, California Institute of Technology, under contract with NASA.


\clearpage


\begin{figure*}[t]

\begin{center}
{\bf FIGURE CAPTIONS}
\end{center}


\figcaption[]{{\it a}. Optical image from Arp (1966) in which Arp~119N is the northern, elliptical member and Arp~119S is the southern, peculiar member.  (North is up and east is to the left in all figures of this paper.) {\it b}. B-band image of Arp~119S (Marziani, et al.\ 1994).  The superimposed contours follow the logarithm of the intensity. The highest contour shown is for an intensity 25\% of the highest intensity seen at this resolution.  The lowest contour is for an intensity 1\% of the highest level.  We have emphasized this density range to facilitate comparison with the models.  {\it c}. H$\alpha$ image of Arp~119S (Marziani et al.\ 1994).  The superimposed contours follow the logarithm of the intensity.  The highest contour shown is for an intensity 85\% of the highest intensity detected at this resolution.  The lowest contour is for an intensity 1\% of the highest value.  {\it d}. CO(1--0) image and contours for Arp~119S (Gao 1996).  The contours were applied by Gao and represent the measured CO(1--0) flux.  The peak contour level is 11.7~Jy~B$^{-1}$~km~s$^{-1}$.  The lowest contour levels are separated by 0.4~Jy~B$^{-1}$~km~s$^{-1}$, the highest levels are separated by 2~Jy~B$^{-1}$~km~s$^{-1}$, and the intermediate levels are separated by about 1.2~Jy~B$^{-1}$~km~s$^{-1}$. \label{fig1}}

\end{figure*}

\begin{figure*}[t]


\figcaption[]{{\it a}. J-Band image obtained using the NICMOS camera on the 1-meter Mt.\ Laguna telescope, with logarithmic contours superimposed that follow equal intensity.  The highest contour shown is for an intensity level only 38\% of the highest value found for the nucleus of Arp~119S at the resolution of these observations.  The smallest intensity contour is at only 2\% of this peak value.  Contours are not shown for the highest intensities because they would be very close together in the Arp~119S nuclear region.  The maximum intensity in the nucleus of Arp~119N is about 2.4 times the highest value for Arp~119S.  {\it b}. The corresponding H-Band image, with contours superimposed as in 2a. The highest contour level is for an intensity 50\% that of the peak intensity detected in the nucleus of Arp~119S.  The lowest contour is for an intensity 6\% that of the peak value.  The maximum intensity seen in Arp~119N is 2.7 times that for Arp~119S. \label{fig2}}

\end{figure*}

\begin{figure*}[t]


\figcaption[]{Here we present information on the history of the simulated collision from the Earth view.  {\it a}.  The initial configuration of the disk galaxy and the elliptical is shown, where the former is represented by its disk, and the center-of-mass of the elliptical is indicated by a sphere.  From this perspective, the lower part of the disk faces the Earth.  The dashed line shows the approximate trajectory of the elliptical with respect to the disk during the collision, with the arrowhead at the upper end showing the direction of the elliptical's motion.  The solid arrow at the edge of the disk indicates the direction of rotation for the simulated disk galaxy.  {\it b}. The simulated disk gas (gray shades) and the elliptical ``star-particles'' (black) are shown for a model taken from the simulation which corresponds to a time of 24~Myr after the collision for the Arp~119 system. The SPH particle positions displayed here are shaded according to the density they represent, with the darker particles marking the highest density.  (The number density of our SPH particles is not an indication of the gas mass density, because the particles do not have equal masses.)  {\it c}. A similar representation of the disk gas and the elliptical galaxy at a time corresponding to 46~Myr after the collision.  {\it d}.  The model representing the disk gas and elliptical at the current epoch, that is, 71~Myr after the collision. \label{fig3}} 

\end{figure*}

\begin{figure*}[t]


\figcaption[]{A representation of the simulated disk stars (gray) and elliptical particles (black) as viewed from the Earth perspective for the model representing the current epoch.  This timestep corresponds to a time 71~Myr after the ``collision'' of the Arp~119 components.  The elliptical particles represent both stars and the ``dark matter.'' \label{fig4}}

\end{figure*}

\begin{figure*}[t]


\figcaption[]{Views of the current epoch disk gas as represented by the SPH particles, shaded by density, with darkest the most dense, for two different orientations. {\it a}. A ``face-on view,'' along the disk's rotation axis.  {\it b}. An ``edge-on'' view in the original plane of the disk along a direction vector that lies in the plane defined by the line-of-sight from Earth and the disk spin axis. \label{fig5}}

\end{figure*}

\begin{figure*}[t]


\figcaption[]{In-plane, mass-averaged radial and azimuthal velocities for the simulated gas disk at the current epoch.  Radial and azimuthal components are defined using the disk galaxy's center of mass as the coordinate origin.  The disk is viewed from the face-on perspective, from the dirction of approach of the elliptical. The impact point is located along the horizontal axis, to the right of the galactic center (see Figure 3a for reference). The disk is rotating counter-clockwise in this view, as it is when viewed from the Earth-perspective.  {\it a}. Azimuthal, in-plane velocities.  Positive velocities are counter-clockwise about the center of mass.  {\it b}. Radial, in-plane velocities.  Positive velocities are outward from the center of mass.  {\it c}. Ratio of the radial in-plane velocities to the azimuthal in-plane velocities.  At each point across the surface of the disk, the magnitude of the radial velocity (Figure 6a) is divided by that of the azimuthal velocity (Figure 6b).  Note that in the outer parts of the disk, the radial expansion velocities can exceed the rotational velocities by a factor of 2 or more. \label{fig6}}

\end{figure*}

\begin{figure*}[t]


\figcaption[]{Average line-of-sight velocity field for the disk gas, viewed from the Earth perspective.  For each location in the field, the velocity is calculated as the mass-weighted velocity of all of the SPH particles in the column along the line of sight. \label{fig7}}

\end{figure*}


\end{document}